\documentclass[10pt,conference]{IEEEtran}
\usepackage{amssymb}
\usepackage{amsmath}
\usepackage{algorithm}
\usepackage{algorithmic}
\usepackage{graphicx}
\begin{document}

\title{Deformed Statistics Free Energy Model for Source Separation using Unsupervised Learning}

\author{
\authorblockN{R. C. Venkatesan}
\authorblockA{Systems Research Corporation\\
Aundh, Pune 411007, India \\
Email: ravi@systemsresearchcorp.com} \and
\authorblockN{A. Plastino}
\authorblockA{IFLP, National University La Plata \\
\& National Research Council (CONICET) \\
C. C., 727 1900, La Plata,Argentina\\
Email: plastino@venus.fisica.unlp.edu.ar} }
 \maketitle

\begin{abstract}
A generalized-statistics variational principle for source
separation is formulated by recourse to  Tsallis' entropy
subjected to the additive duality and employing constraints described by normal averages.  The
variational principle is amalgamated with Hopfield-like learning
rules resulting in an unsupervised learning model.  The update
rules are formulated with the aid of $q$-deformed calculus.
Numerical examples exemplify the efficacy of this model.

\end{abstract}

\section{Introduction}
Recent studies have suggested that minimization of the Helmholtz
free energy in statistical physics [1] plays a central role in
understanding action, perception, and learning (see [2] and the
references therein).  In fact, it has been suggested that the
principle of free energy minimization is even more fundamental
than the redundancy reduction principle (also known as the
principle of efficient coding) articulated by Barlow [3] and later
formalized by Linsker as the Infomax principle [4]. Specifically,
the principle of efficient coding states that the brain should
optimize the mutual information between its sensory signals and
some parsimonious neuronal representations. This is identical to
optimizing the parameters of a generative model to maximize the
accuracy of predictions, under complexity constraints. Both are
mandated by the free-energy principle, which can be regarded as a
probabilistic generalization of the Infomax principle.

The Infomax principle has been central to the development of
independent component analysis (ICA) and the allied problem of
blind source separation (BSS) [5].  Within the ICA/BSS context,
very few models based on minimization of the free energy exist,
the most prominent of them originated by Szu and co-workers (eg.
see Refs. [6,7]) to achieve source separation in remote sensing
(i.e. hyperspectral imaging (HSI)) using the maximum entropy
principle. The ICA/BSS problem may be summarized in terms of  the
relation
\begin{equation}
\textbf{A}\textbf{s}=\textbf{x},
\end{equation}
where $\textbf{s}$ is the \textit{unknown} source vector to be
extracted, $\textbf{A}$ is the \textit{unknown} mixing matrix
(also known as reflectance matrix or material abundance matrix in
HSI), and $\textbf{x}$ is the \textit{known} vector of observed
data. The Helmholtz free energy is described within the framework
of Boltzmann-Gibbs-Shannon (B-G-S) statistics as
\begin{equation}
F(T)=U-k_BTS,
\end{equation}
where $T$ is the thermodynamic temperature (or haemostatic
temperature in the parlance of cybernetics), $k_B$  the Boltzmann
constant, $U$  the internal energy, and $S$  Shannon's entropy. A
more principled and systematic manner in which to study free
energy minimization within the context of the maximum entropy
principle (MaxEnt) is by substituting the minimization of the
Helmholtz free energy principle with the maximizing of the Massieu
potential [8]
\begin{equation}
\Phi(\beta)=S-\beta U,
\end{equation}
where $ \beta  = \frac{1}{{k_B T}}$ is the inverse thermodynamic
temperature.  The Massieu potential is the Legendre transform of
the Helmholtz free energy, i.e.: $ \Phi \left( \beta  \right) =  -
\frac{{F\left( T \right)}}{T}$.

The generalized (also, interchangeably, nonadditive, deformed, or
nonextensive) statistics of Tsallis has recently been the focus of
much attention in statistical physics, complex systems, and allied
disciplines [9].  Nonadditive statistics suitably generalizes the
extensive, orthodox B-G-S one. The scope of Tsallis statistics has
lately been extended to studies in lossy data compression in
communication theory [10] and machine learning [11,12].

It is important to note that power law distributions like the
q-Gaussian distribution cannot be accurately modeled within the
B-G-S framework [9]. One of the most
 commonly encountered source of q-Gaussian distributions occurs in
 the process of normalization of measurement data using
 \textit{Studentization} techniques [13].  q-Gaussian behavior is
 also exhibited by elliptically invariant data, which
 generalize spherically symmetric distributions. q-Gaussian's are also
 an excellent approximation to correlated Gaussian data, and other
 important and fundamental physical and biological processes (for example, see [14] and the references therein).

This paper intends to accomplish the following objectives:
\begin{itemize} \item $(i)$ to formulate and solve a variational principle for source separation using
the maximum dual Tsallis entropy with constraints defined by
normal averages expectations, \item $(ii)$ to amalgamate the
variational principle with Hopfield-like learning rules [15] to
acquire information regarding unknown parameters via an
unsupervised learning paradigm,
\item$(iii)$ to formulate a numerical framework for the
generalized statistics unsupervised learning model and demonstrate,
with the aid of numerical examples for separation of independent
sources (\textit{endmembers}), the superiority of the generalized
statistics source separation model \textit{vis-\'{a}-vis} an
equivalent B-G-S model for a single pixel.\
\end{itemize}

It is important to note that by amalgamating the
information-theoretic model with the Hopfield model, $[\textbf{A}]$
acquires the role of the Associative Memory (AM) matrix.
\textit{Further, employing a Hopfield-like learning rule renders the
model presented in this paper readily amenable to hardware
implementation using Field Programmable Gate Arrays (FPGA's)}.

The additive duality is a fundamental property in generalized
statistics [9]. One implication of the additive duality is that it
permits a deformed logarithm defined by a given nonadditivity
parameter (say, $q$) to be inferred from its \textit{dual deformed}
logarithm parameterized by: $q^*=2-q$.  This paper derives a
variational principle for source separation using the dual Tsallis
entropy using normal averages constraints.  This approach has been
previously utilized (for eg. Ref. [16]), and possess the property of seamlessly yielding a $q^*$-deformed exponential form on variational extremization.

An important issue to address concerns the manner in which
expectation values are computed.
 Of the various forms in which expectations may be defined in nonextensive statistics has,
 only the linear constraints originally employed by Tsallis [9] (also
known as \textit{normal averages}) of the form: $ \left\langle A
\right\rangle = \sum\limits_i {p_i } A_i $, has been found to be
physically satisfactory and consistent with both the generalized
H-theorem and the generalized \textit{Stosszahlansatz} (molecular
chaos hypothesis) [17, 18]. A re-formulation of the variational
perturbation approximations in nonextensive statistical physics
followed [18], via an application of $q$-deformed calculus [19].
Results from the study in Ref. [19] have been successfully
utilized in Section IV of this paper.

This introductory Section is concluded by briefly describing the
suitability of employing a generalized statistics model to study the
source separation problem.  First, in the case of remote sensing
applications, and even more so in the case of HSI, the observed data
 are highly correlated, even in the case of a single pixel.  Next, the
observed data are required to be normalized (scaled).  The
\textit{Studentization} process is one of the most prominent
methods utilized to normalize the observed data [20,21].  Both
these features lead to an excursion from the Gaussian framework
(B-G-S statistics) and result in q-Gaussian pdf's characterized by
the $q$-deformed exponential: $exp_q(-x)=\left[ {1 - \left( {1 -
q} \right)x} \right]^{\frac{1}{{1 - q}}}$, which maximizes the
Tsallis entropy.

\section{Theoretical preliminaries}

The Section introduces the essential concepts around which this
communication revolves. The Tsallis entropy is defined as [9]
\begin{equation}
\begin{array}{l}
S_q \left( X \right) =  - \sum\limits_x {p\left( x \right)} ^q \ln
_q p\left( x \right).\\
\end{array}
\end{equation}
The \textit{q-deformed} logarithm and the \textit{q-deformed}
exponential are defined as [9, 19]
\begin{equation}
\begin{array}{l}
\ln _q \left( x \right) = \frac{{x^{1 - q} - 1}}{{1 - q}}, \\
and, \\
\exp _q \left( x \right) = \left\{ \begin{array}{l}
 \left[ {1 + \left( {1 - q} \right)x} \right]^{\frac{1}{{1 - q}}} ;1 + \left( {1 - q} \right)x \ge 0, \\
 0;otherwise \\
 \end{array} \right.
\end{array}
\end{equation}
Note that as $q\rightarrow 1$, (4) acquires the form of the
equivalent B-G-S entropies.  Likewise in (5), $\ln_q(x)\rightarrow
\ln(x)$ and $\exp_q(x)\rightarrow \exp(x)$.  The operations of
\textit{q-deformed} relations are governed by \textit{q-deformed}
algebra and \textit{q-deformed} calculus [19]. Apart from providing
an analogy to equivalent expressions derived from B-G-S statistics,
\textit{q-deformed} algebra and \textit{q-deformed} calculus  endow
generalized statistics with a unique information geometric
structure. The \textit{q-deformed} addition $ \oplus_q $ and the
\textit{q-deformed} subtraction $ \ominus_q $ are defined as [19]
\begin{equation}
\begin{array}{l}
 x \oplus_q y = x + y + \left( {1 - q} \right)xy, \\
  \ominus_q y = \frac{{ - y}}{{1 + \left( {1 - q} \right)y}};1+(1-q)y > 0 \\
  \Rightarrow x \ominus_q y = \frac{{ x- y}}{{1 + \left( {1 - q} \right)y}} \\
 \end{array}
\end{equation}

The \textit{q-deformed} derivative, is defined as [19]
\begin{equation}
D^x_{q } F\left( x  \right) = \mathop {\lim }\limits_{y \to x }
\frac{{F\left( x  \right) - F\left( y \right)}}{{x \ominus_q y }} =
\left[ {1 + \left( {1 - q} \right)x } \right]\frac{{dF\left( x
\right)}}{{dx }}
\end{equation}
As $q\rightarrow1$, $D^x_{q } F\left( x \right)\rightarrow
dF(x)/dx$, the Newtonian derivative. The Leibnitz rule for
\emph{deformed} derivatives [19] is
\begin{equation}
D_{q}^x  \left[ {A\left( x \right)B\left( x \right)} \right] =
B\left( x \right)D_{q}^x A\left( x \right) + A\left( x
\right)D_{q}^x B\left( x \right).
 \end{equation}

 Re-parameterizing (5) via the
\textit{additive duality} [10]: $q^*=2-q $, yields the \textit{dual
deformed} logarithm and exponential
\begin{equation}
\begin{array}{l}
 \ln _{q^*}  \left( x \right) =  - \ln _q  \left( {\frac{1}{x}} \right), and, \exp _{q^*}  \left( x \right) = \frac{1}{{\exp _q  \left( { - x} \right)}}. \\
 \end{array}
\end{equation}

The dual Tsallis entropy is defined by [10, 16]
\begin{equation}
S_{q^*} \left( X \right) =  - \sum\limits_x {p\left( x \right)\ln
_{q^*} } p\left( x \right).
\end{equation}

Here, $  \ln_{q^*}(x) = \frac{{x^{1 - q^*} - 1}}{{1 - q^*}} $.
\textit{The dual Tsallis entropies acquire a form similar to the
B-G-S entropies, with $ \ln_{q^*}(\bullet) $ replacing $
\ln(\bullet) $}.

\section{Variational principle}

Consider the Lagrangian
\begin{equation}
\begin{array}{l}
\Phi _{q^ *  } [ {s_j }] =  - \sum\limits_j {s_j \ln _{q^ *  } s_j }  -
\sum\limits_{i = 1}^N {\sum\limits_{j = 1}^N {\lambda _i } \left( {A_{ij} s_j  - x_i } \right)}  \\
  +  {\lambda _0}\left( {\sum\limits_{j = 1}^N {s_j }  - 1} \right), \\
 \end{array}
\end{equation}
subject to the component-wise constraints
\begin{equation}
\sum\limits_{j = 1}^N {s_j }  = 1,\,\,\,{\rm and}\,\,\,
\sum\limits_{j = 1}^N {A_{ij} s_j  = x_i }.
\end{equation}
Clearly, the RHS of the Lagrangian (11) is the $q^*$-deformed
Massieu potential: $\Phi_{q^*}[\lambda]$, subject to the
normalization constraint on $s_j$. The variational extremization of
(11), performed using the Ferri-Martinez-Plastino methodology [22],
 leads to
\begin{equation}
\begin{array}{l}
  \Rightarrow  - \frac{{\left( {2 - q^ *  } \right)}}{{\left( {1 - q^ *  } \right)}}s_j^{1 - q^ *  }  - \sum\limits_{i = 1}^N {\lambda _i A_{ij} }  + \lambda _0  = 0 \\
  \Rightarrow s_j  = \left[ {\frac{{\left( {1 - q^ *  } \right)}}{{\left( {q^ *   - 2} \right)}}\left( -{\lambda _0  + \sum\limits_{i = 1}^N {\lambda _i A_{ij} } } \right)} \right]^{\frac{1}{{1 - q^ *  }}}  \\
\end{array}
\end{equation}

Multiplying the second relation in (13) by $s_j$ and summing over
all $j$, yields after application of the normalization condition in
(12)
\begin{equation}
 - \frac{{\left( {2 - q^ *  } \right)}}{{\left( {1 - q^ *  } \right)}}\aleph _{q^ *  }  - \sum\limits_{j = 1}^N {\sum\limits_{i = 1}^N {\lambda _i A_{ij} s_j } }  = -\lambda
 _0,
\end{equation}
where: $ \aleph _{q^ *  }  = \sum\limits_{j = 1}^N {s_j^{2 - q^ * }
}$, and substituting (14) into the third relation in (13) yields
\begin{equation}
\begin{array}{l}
 s_j  \\
  = \left[ {\aleph _{q^ *  }  + \left( {1 - q^ *  } \right)\sum\limits_{j = 1}^N {\sum\limits_{i = 1}^N {\tilde \lambda _i A_{ij} s_j } }  - \left( {1 - q^ *  } \right)\sum\limits_{i = 1}^N {\tilde \lambda _i A_{ij} } } \right]^{\frac{1}{{1 - q^ *  }}} ; \\
 \tilde \lambda _i  = \frac{{\lambda _i }}{{\left( {2 - q^ *  } \right)}}. \\
 \end{array}
\end{equation}
Eq. (15) yields after some algebra
\begin{equation}
\begin{array}{l}
 s_j   = \frac{{\exp _{q^ *  } \left( { - \sum\limits_{i = 1}^N {\tilde \lambda _i^ *  A_{ij} } } \right)}}{{\left( {\aleph _{q^ *  }  + \left( {1 - q^ *  } \right)\sum\limits_{j = 1}^N {\sum\limits_{i = 1}^N {\tilde \lambda _i A_{ij} s_j } } } \right)^{^{\frac{1}{{q^ *   - 1}}} } }}, \\
 \end{array}
\end{equation}
where
\begin{equation}
\begin{array}{l}
\tilde \lambda _i^ *   = \frac{{\tilde \lambda _i }}{{\aleph _{q^ *  }  + \left( {1 - q^ *  } \right)\sum\limits_{j = 1}^N {\sum\limits_{i = 1}^N {\tilde \lambda _i A_{ij} s_j } } }}, \\
 {\rm and}, \\
 \left( {\aleph _{q^ *  }  + \left( {1 - q^ *  } \right)\sum\limits_{j = 1}^N {\sum\limits_{i = 1}^N {\tilde \lambda _i A_{ij} s_j } } } \right)^{^{\frac{1}{{q^ *   - 1}}} }  = \tilde Z_{q^ *  } . \\
 \end{array}
\end{equation}
Here $\tilde Z_{q^*}$ is the canonical partition function, where: $
\tilde Z_{q^*}  = \sum\limits_{j = 1}^N {\exp _{q^ *  } \left( { -
\sum\limits_{i = 1}^N {\tilde \lambda _i^ *  A_{ij} } } \right)} $.
The dual Tsallis entropy takes the form
\begin{equation}
\begin{array}{l}
 S_{q^ *  } [ {s }] = \frac{{\aleph _{q^ *  }  - 1}}{{\left( {q^ *   - 1} \right)}};
\sum\limits_{j = 1}^N {s_j }  = 1 \\
  \Rightarrow \aleph _{q^ *  }  = 1 + \left( {q^ *   - 1} \right)S_{q^ *  } [ {s }] \\
 \end{array}
\end{equation}
Substituting now (18) into the expression for: $\tilde Z_{q^*}$ in
(17) results in
\begin{equation}
\begin{array}{l}
 - \ln _{q^ *  } \left( {\frac{1}{{\tilde Z_{q^ *  } }}} \right) = S_{q^ *  } [ {s}] - \sum\limits_{j = 1}^N {\sum\limits_{i = 1}^N {\tilde \lambda _i A_{ij} s_j } }  = \Phi_{q^*} \left[ {\tilde \lambda } \right]. \\
 \end{array}
\end{equation}
Clearly, $ \Phi_{q^*} \left[ {\tilde \lambda } \right]$ in (19) is
a $q^*$-deformed Massieu potential.  By substituting (18) into
(14) we arrive at
\begin{equation}
\begin{array}{l}
 S_{q^ *  } \left[ s \right] - \sum\limits_{j = 1}^N {\sum\limits_{i = 1}^N {\tilde \lambda _i A_{ij} s_j } }  = -\tilde \lambda _0 + \frac{1}{{\left( {1 - q^ *  } \right)}} = \hat \lambda _0 ; \\
 \tilde \lambda _0  = \frac{\lambda_0 }{{\left( {2 - q^ *  } \right)}}. \\
 \end{array}
\end{equation}
Again, $\hat \lambda_0$ in (20) is a $q^*$-deformed Massieu
potential: $\Phi_{q^*}[\tilde\lambda]$. We wish to relate
$\hat\lambda_0$ and $\tilde Z_{q^*}$. To this end, comparison of
(19) and (20) yields
\begin{equation}
\begin{array}{l}
 \hat\lambda_0=-\tilde \lambda _0  + \frac{1}{{\left( {1 - q^ *  } \right)}} =  - \frac{{\tilde Z_{q^*}^{q^ *   - 1} }}{{\left( {1 - q^ *  } \right)}} + \frac{1}{{\left( {1 - q^ *  } \right)}} \\
  \Rightarrow \tilde Z_{q^*}  = \left[ {\left( {1-q^ *  } \right)
  \tilde \lambda _0 } \right]^{\frac{1}{{q^ *   - 1}}} ;
  \tilde \lambda _0  = \frac{{\lambda _0 }}{{\left( {2 - q^ *  } \right)}}, \\
 \end{array}
\end{equation}
so that, by substituting (18) into (15) and then invoking (20) we
get
\begin{equation}
\begin{array}{l}
 s_j  = \left[ {1 - \left( {1 - q^ *  } \right)\left( {\sum\limits_{i = 1}^N {\tilde \lambda _i A_{ij} }  + \hat \lambda _0 } \right)} \right]^{\frac{1}{{1 - q^ *  }}} ; \\
 \hat \lambda _0  =  - \tilde \lambda _0  + \frac{1}{{\left( {1 - q^ *  } \right)}}. \\
 \end{array}
\end{equation}
Here, (22) is re-defined with the aid of (20) as
\begin{equation}
\begin{array}{l}
 s_j  = \frac{{\left[ {1 - \left( {1 - q^ *  } \right)\sum\limits_{i = 1}^N {\tilde \lambda _i^ *  A_{ij} } } \right]^{\frac{1}{{1 - q^ *  }}} }}{{\left[ {1 - \left( {1 - q^ *  } \right)\hat \lambda _0 } \right]^{\frac{1}{{q^ *   - 1}}} }}  = \frac{{\left[ {1 - \left( {1 - q^ *  } \right)\sum\limits_{i = 1}^N {\tilde \lambda _i^ *  A_{ij} } } \right]^{\frac{1}{{1 - q^ *  }}} }}{{\tilde Z_{q^ *  } }}; \\
 {\rm where} \\
\tilde \lambda _i^ *   = \frac{{\tilde \lambda _i }}{{1 - \left( {1
- q^ *  } \right)\hat \lambda _0 }}, \tilde Z_{q^ * } =
\sum\limits_{j = 1}^N {\left[ {1 - \left( {1 - q^ *  } \right)
\sum\limits_{i = 1}^N {\tilde \lambda _i^ *  A_{ij} } } \right]^{\frac{1}{{1 - q^ *  }}} }.  \\
 \end{array}
\end{equation}
 With the aid of
(21), (22) is re-cast in the form
\begin{equation}
\begin{array}{l}
s_j  = \frac{{\exp _{q^ *  } \left( { - \sum\limits_{i = 1}^N
{\tilde \lambda _i^ *  A_{ij} } } \right)}}{{\left[ {\left( {1 - q^
*  } \right)\tilde \lambda _0 } \right]^{\frac{1}{{q^ *   - 1}}}
}};\\
\,{\rm where},\,\, \tilde \lambda _i  = \frac{{\lambda _i }}{{\left(
{2 - q^ *  } \right)}},\tilde \lambda _0  = \frac{{\lambda _0
}}{{\left( {2 - q^ *  } \right)}},
\tilde \lambda _i^ *   = \frac{{\tilde \lambda_i }}{{\left[ {\left( {1 - q^ *  } \right)\tilde \lambda _0 } \right]}}. \\
\end{array}
\end{equation}
Finally, invoking the normalization of $s_j$, (24) yields
\begin{equation}
\left[ {\left( {1 - q^ *  } \right)\tilde \lambda _0 }
\right]^{\frac{1}{{q^ *   - 1}}}  = \sum\limits_{j = 1}^N {\left[ {1
- \left( {1 - q^ *  } \right)\sum\limits_{i = 1}^N {\tilde \lambda
_i^ *  A_{ij} } } \right]^{\frac{1}{{1 - q^ *  }}} .}
\end{equation}
Note the \textit{self-referential} nature of (23) in the sense
that: $\tilde\lambda_{i}^*$ (defined in (20) and (23) is a function of
$\tilde\lambda_0$. The Lagrange multiplier $\tilde\lambda_{i}^*$
is henceforth defined in this paper as the \textit{dual normalized
Lagrange force multiplier}.

\section{Unsupervised Learning rules}

The process of unsupervised learning is amalgamated to the above
information theoretic structure via a Hopfield-like learning rule
to update the AM matrix $[\textbf{A}]$ in the case of  a
perturbation $\Delta x_j$ of the observed data
\begin{equation}
\begin{array}{l}
 \frac{{dx_j }}{{dt}} =  \frac{{\partial \tilde \Phi _{q^ *  }^ *  \left[ {s_j } \right]}}{{\partial s_j }} \\
  = -\frac{{1 - \left( {1 - q^
*  } \right)\tilde \lambda }}{{\left( {1 - q^ *  } \right)\tilde
\lambda _0 }} - \frac{{\ln _{q^ *  } s_j }}{{\tilde \lambda _0 }} -
\left( {1 - q^ *  } \right)\sum\limits_{i = 1}^N {\tilde \lambda _i^
*  } A_{ij} \\
  \Rightarrow \Delta x_j  \\
  =
-\left[ {\frac{{1 - \left( {1 - q^ *  } \right)\tilde \lambda _0
}}{{\left( {1 - q^ *  } \right)\tilde \lambda _0 }} + \frac{{\ln
_{q^ *  } s_j }}{{\tilde \lambda _0 }} + \left( {1 - q^
*  } \right)\sum\limits_{i = 1}^N {\tilde \lambda _i^{ * } } A_{ij} } \right]\Delta t; \\
where, \tilde \Phi _{q^ *  }^ *  \left[ {s_j } \right] = \frac{{\Phi _{q *
} \left[ {s_j } \right]}}{{\left( {2 - q^ *  } \right)\tilde \lambda
_0 }},
\end{array}
\end{equation}
which is obtained from the first relation in (13) and (24). Gradient
ascent along with (24) originates the second learning rule
\begin{equation}
\begin{array}{l}
 \frac{{dx_j }}{{dt}} =   \frac{{\partial \Phi _{q^ *  }^ *  \left[ {s_j } \right]}}{{\partial A_{ij} }} = -\tilde \lambda _i^ *  s_j  \Rightarrow \Delta x_j  = -\left( {\tilde \lambda _i^ *  s_j } \right)\Delta t; \\
where, \Phi _{q^ *  }^ *  \left[ {s_j } \right] = \frac{{\Phi _{q^ *  } \left[ {s_j } \right]}}{{ \left( {1 - q^ *  } \right)\tilde \lambda _0}}. \\
 \end{array}
\end{equation}
In (26) and (27), $\Phi _{q^ *  } \left[ {s_j } \right]$ is
the LHS of the Lagrangian (11).

Now, a \textit{critical} update rule is that for the change in the
\textit{dual normalized Lagrange force multipliers }
$\tilde\lambda_i^*$ resulting from a perturbation $\Delta x_j$ in
the observed data.  Usually (as stated within the context of the
B-G-S framework), such an update would entail a Taylor-expansion
yielding up to the first order: $ \Delta x_j  = \sum\limits_{k =
1}^N {\frac{{\partial x_j }}{{\partial \tilde \lambda _k^ *  }}}
\Delta \tilde \lambda _k^ *$. Such an analysis is valid only for
distributions characterized by the regular exponential $exp(-x)$.
For probability distributions characterized by $q$-deformed
exponentials, i.e., the ones we face here, such a perturbation
treatment would lead to un-physical results [18].

Thus, following the prescription given in Ref. [18], for a function:
$ F\left( \tau \right) = \sum\limits_n {F\left( {\tau _n } \right)}
$ the chain rule yields: $ \frac{{dF \left( \tau \right)}}{{d\tilde
\lambda _k^ * }} = \frac{{dF \left( \tau \right)}}{{d\tau
}}\frac{{d\tau }}{{d\tilde \lambda _k^ * }} \ $.
 Thus, replacing the Newtonian derivative: $ \frac{{dF \left( \tau
\right)}}{{d\tau }} $ by the \textit{$q^*$-deformed} one defined by
(7) (see Ref. [19]): $ D^\tau_{q^*}F(\tau) = \left[ {1 + \left( {1 -
q^*} \right)\tau } \right]\frac{{dF\left( \tau \right)}}{{d\tau }} $
and defining:  $ D_{q^* }^\tau F \left( \tau \right)\frac{{d\tau
}}{{d\tilde \lambda _k^ *}} =\delta _{q^*,\tau} F \left( \tau
\right) $ as well, facilitates the desired transformation: $
\frac{{dF \left( \tau \right)}}{{d\tilde \lambda _k^ * }} \to \delta
_{ q^*,\tau} F \left( \tau \right) $.
 Consequently, the update rule for $\tilde \lambda _k^ *$ is re-formulated
via $q$-deformed calculus in the fashion
\begin{equation}
\Delta x_j  = \sum\limits_{k = 1}^N {\left[ {D_{q^ *  }^\tau
\sum\limits_{i = 1}^N {A_{ji} s_i } } \right]} \Delta \tilde \lambda
_k^ *   = \sum\limits_{k = 1}^N {\left[ {\sum\limits_{i = 1}^N
{D_{q^ *  }^\tau  A_{ji} s_i } } \right]} \Delta \tilde \lambda _k^
*.
\end{equation}
Additionally, setting: $
 - A_{ik} \tilde \lambda _k^ *   = \tau $ in (23) leads to
\begin{equation}
s_j  = \frac{{\left[ {1 + \left( {1 - q^ *  } \right)\tau }
\right]^{\frac{1}{{1 - q^ *  }}} }}{{\tilde Z_{q^ *  } }}.
\end{equation}
Employing at this stage the Leibnitz rule for $q^*$-deformed
derivatives (and replacing $q$ by $q^*$ in (8)), the term within
square parenthesis RHS in (28) yields
\begin{equation}
\begin{array}{l}
 \sum\limits_{i = 1}^N {D_{q^ *  }^\tau  A_{ji} s_i }  = \sum\limits_{i = 1}^N {\left\{ {\frac{{A_{ji} }}{{\tilde Z_{q^ *  } }}D_{q^ *  }^\tau  \left[ {1 + \left( {1 - q^ *  } \right)\tau } \right]^{\frac{1}{{1 - q^ *  }}} } \right.}  \\
 \left. { + A_{ji}
 \left[ {1 + \left( {1 - q^ *  } \right)\tau } \right]^{\frac{1}{{1 - q^ *  }}} D_{q^ *  }^\tau
 \left( {\frac{1}{{\tilde Z_{q^ *  } }}} \right)} \right\}, \\
 \end{array}
\end{equation}
a relation that, after expansion turns into
\begin{equation}
\begin{array}{l}
 \sum\limits_{i = 1}^N {D_{q^ *  }^\tau  A_{ji} s_i } \\
  = \sum\limits_{i = 1}^N {\left\{ {\frac{{A_{ji} }}{{\tilde Z_{q^ *  } }}\left[ {1 + \left( {1 - q^ *  } \right)\tau } \right]\frac{{\partial \tau }}{{\partial \tilde \lambda _k^ *  }}\frac{\partial }{{\partial \tau }}\left[ {1 + \left( {1 - q^ *  } \right)\tau } \right]^{\frac{1}{{1 - q^ *  }}} } \right.}  \\
 \left. { + A_{ji} \left[ {1 + \left( {1 - q^ *  } \right)\tau } \right]^{\frac{1}{{1 - q^ *  }}} D_{q^ *  }^\tau  \left( {\frac{1}{{\tilde Z_{q^ *  } }}} \right)} \right\} \\
  = \sum\limits_{i = 1}^N {\left\{ { - \frac{{A_{ji} }}{{\tilde Z_{q^ *  } }}\left[ {1 + \left( {1 - q^ *  } \right)\tau } \right]^{\frac{1}{{1 - q^ *  }}} A_{ik} } \right.}  \\
 \left. { - A_{ji} \left[ {1 + \left( {1 - q^ *  } \right)\tau } \right]^{\frac{1}{{1 - q^ *  }}} \left[ {1 + \left( {1 - q^ *  } \right)\tau } \right]\frac{{\partial \tau }}{{\partial \tilde \lambda _k^ *  }}\tilde Z_{q^ *  }^{ - 2} \frac{{\partial \tilde Z_{q^ *  } }}{{\partial \tau }}} \right\} \\
  =  - \sum\limits_{i = 1}^N {A_{ji} s_i A_{ik} }  \\
  + \sum\limits_{i = 1}^N {A_{ji} \frac{{\left[ {1 + \left( {1 - q^ *  } \right)\tau } \right]^{\frac{1}{{1 - q^ *  }}} }}{{\tilde Z_{q^ *  } }}} \sum\limits_{k = 1}^N {\frac{{A_{ik} }}{{\tilde Z_{q^ *  } }}} \left[ {1 + \left( {1 - q^ *  } \right)\tau } \right]^{\frac{1}{{1 - q^ *  }}}  \\
  =  - \sum\limits_{i = 1}^N {A_{ji} s_i A_{ik} }  + x_j x_k . \\
 \end{array}
\end{equation}
Finally, the update rule for $\tilde \lambda _k^ *$ with respect
to $\Delta x_j$ adopts the appearance
\begin{equation} \Delta x_j
= \sum\limits_{k = 1}^N {\left( {x_j x_k  - \sum\limits_{i = 1}^N
{A_{ji} s_i A_{ik} } } \right)\Delta \tilde \lambda _k^ *  }.
\end{equation}

\section{Numerical computations}
The procedure for our double recursion problem is summarized in
the pseudo-code below

 \begin{algorithm}
 \caption{Generalized Statistics Source Separation Model}
 \begin{algorithmic}
\STATE $(1.)$~\textbf{Input}: $(i)$. Observed data:
$\textbf{x}$,~$(ii)$. Trial values of dual normalized Lagrange
force multipliers: $\tilde\lambda^*$,~$(iii)$. Dual nonadditive
parameter: $q^*$.

\STATE $(2.)$~\textbf{Initialization}:\\
Obtain $A_{ij}^{(0)}$ from:$A_{ij}^{(0)}=x_i\sigma_{q^*}(x_j)$+ 50
$\%$ random noise to break any rank-1 singularity. The
$q^*$-deformed sigmoid logistic function is: $\sigma_{q^*}(x_j) =
\frac{1}{{1 + \exp _{q^
* } \left( { - x_i } \right)}}$.

 \STATE $(3.)$~\textbf{First Recursion} \\

$(i)$ Compute: $\tilde Z_{q^*}^{(0)}$ from (23), \\
$(ii)$ Compute:$\hat\lambda^{(0)}$ from (21),\\
$(iii)$ Compute: $s_j^{(0)}$, $\tilde\lambda_i^{(0)}$, and $\tilde\lambda_0^{(0)}$ from (23)/(24), \\
$(iv)$ Compute: $x_i^{(0)}$ from (5), thus: $\Delta x_j^{(0)}=x_j^{Known}-x_j^{(0)}$, \\
$(v)$ Compute $\Delta \tilde\lambda_k^{*(0)}$ by inverting (32), \\
$(vi)$ Compute next estimate: $\tilde\lambda_k^{*(1)}=\tilde\lambda_k^{*(0)}+\Delta \tilde\lambda_k^{*(0)}$ . \\

\STATE $(4.)$ \textbf{Second Recursion}\\
$(vii)$  Compute improved estimate of : $A_{ij}^{(1)}$ from (26) by
setting $\Delta t=1$ and solving :$\Delta x_j=- \left[ {\frac{{1 -
\left( {1 - q^ *  } \right)\tilde \lambda _0^{\left( 0 \right)}
}}{{\left( {1 - q^ *  } \right)\tilde \lambda _0^{\left( 0 \right)}
}} + \frac{{\ln _{q^ *  } s_j^{\left( 0 \right)} }}{{\tilde \lambda
_0^{\left( 0 \right)} }} + \left( {1 - q^ *  } \right)\sum\limits_{i
= 1}^N {\tilde \lambda _i^{ * \left( 1
\right)} } A_{ij}^{\left( 1 \right)} } \right]$. \\

\STATE $(5.)$ Go to $(3.)$
\end{algorithmic}
\end{algorithm}

Following the procedure outlined in the above pseudo-code, values of
$\tilde\lambda^*=[0.6228, 0.6337, 0.4577, 0.1095, 0.7252,0.01752,
0.4128]$ and $\textbf{x}= [0.5382, 0.1023, 0.6404, 0.4358,
0.0278,0.2425, 0.3299]$ are provided.  These values are the same as
those in Ref. [7] and constitute experimentally obtained Landsat
data for a single pixel. The difference between the generalized
statistics model presented in this paper and the B-G-S model of [6,7]
lies in the fact that the former has initial inputs of
$\tilde\lambda_i^*$'s, whereas the latter merely has initial inputs of
$\lambda$'s (a far simpler case). \textit{The self-rerentiality in (23) mandates use of $\tilde\lambda_i^*$'s as the primary operational Lagrange multiplier}.  Note that the correlation
coefficient of $x^{Known}$ is unity, a signature of highly
correlated data.  A value of $q^*=0.75$ is chosen. Figure 1 and
Figure 2 depict, vs. the number of iterations,  the source
separation for the generalized statistics model and for the B-G-S
model, respectively. Values of $\textbf{x}$ are denoted by "o"'s. It
is readily appreciated that the generalized statistics exhibits a
more pronounced source separation than the B-G-S model. Owing to the
highly correlated nature of the observed data, such results are to
be expected.

\section{Summary and discussion}
A generalized statistics model for source separation that employs an
unsupervised learning paradigm has been presented in this
communication. This model is shown to exhibit superior separation
performance as compared to an equivalent model derived within the
B-G-S framework.  Our encouraging results should inspire
future work studies on the implications of first-order and
second-order phase transitions of the Massieu potential.  One would
wish for a self-consistent scheme enabling one to obtain
self-consistent values of Lagrange multipliers based on the
principle of phase transitions and symmetry breaking.

\begin{figure}[thpb]
\centering
\begin{center}
\includegraphics[scale=.50]{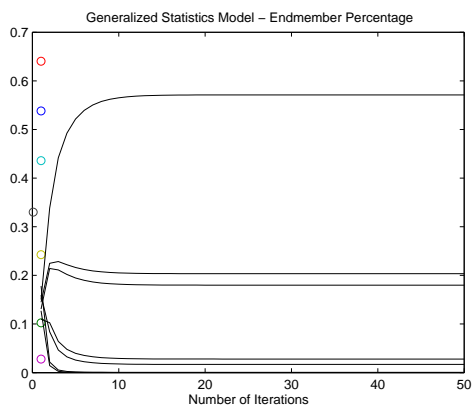}
\caption [loftitle]{Source separation for generalized statistics model  }
\end{center}
\end{figure}

\begin{figure}[thpb]
\centering
\begin{center}
\includegraphics[scale=.500]{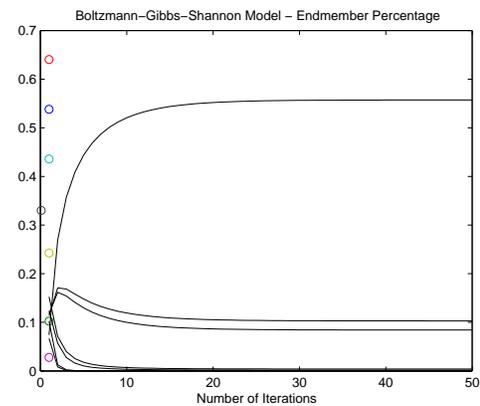}
\caption [loftitle]{Source separation for Boltzmann-Gibbs-Shannon
model }
\end{center}
\end{figure}

\section*{Acknowledgment}
RCV gratefully acknowledges support from \textit{RAND-MSR} contract
\textit{CSM-DI $ \ \& $ S-QIT-101155-03-2009}.



%

\end{document}